# Thermosolutal Marangoni instability in a viscoelastic liquid film: Effect of heating from the free surface

**Rajkumar Sarma and Pranab Kumar Mondal**†

Department of Mechanical Engineering, Indian Institute of Technology Guwahati, Assam, India – 781039

We study the Marangoni instability in a thin polymeric liquid film heated from the free surface. Polymeric solutions are usually the binary mixture of a Newtonian solvent with polymeric solute and exhibit viscoelastic behaviour. In the presence of a temperature gradient, stratification of these polymeric solutes can take place via the Soret effect, which may give rise to the solutocapillary effect at the free surface. Considering the Soret effect and incorporating the effects of gravity, here we analyse the stability characteristics of this polymeric thin film bounded between its deformable free surface and a poorly conductive rigid substrate from below. A linear stability analysis around the quiescent base state reveals that under the combined influences of thermo-solutocapillarity, apart from the monotonic disturbance, two different oscillatory instabilities can emerge in the system. The characteristics of each instability mode are discussed, and a complete stability picture is perceived in terms of the phase diagrams, identifying the model parameter space wherein a particular instability mode can get dominant.

## 1. Introduction

On the free surface of a pure liquid (or a liquid mixture), a sufficiently strong local variation in temperature (or concentration) can lead to the development of surface shear stresses via the thermocapillary (solutocapillary) effect. Such surface stresses have the capability to induce motion in the bulk phase of a small scale system *viz.*, the thin-films, droplets, vapour bubbles, liquid bridges, etc., typically known as the Marangoni convection. The ensuing flow emerges with the formation of beautiful surface patterns that was famously observed by Bénard (1901), Vanhook *et al.* (1997) for pure liquids and subsequently by Zhang *et al.* (2007) and Toussaint *et al.* (2008) for binary liquid mixtures. Marangoni convection is an important area of research due to its emergence in numerous physical and engineering applications including the drying of colloidal films (Yiantsios & Higgins 2006), crystal growth (Boggon *et al.* 1998), laser cladding (Kumar & Roy 2009), fusion welding (Mills *et al.* 1998) and the patterning of liquid metals/polymer films (Arshad *et al.* 2014). Due to the involvement of surface effects rather than the volumetric ones, this convection phenomenon is of crucial importance for the microgravity environment.

† Email address for correspondence: mail2pranab@gmail.com    1

Despite remarkable advancements toward understanding the Marangoni convection for pure/binary Newtonian fluids (for a comprehensive description, see Colinet *et al.* 2001; Shklyaev & Nepomnyashchy 2017), this instability phenomenon in the context of viscoelastic fluids has remained much unexplored. The viscoelastic fluids, e.g. the polymeric solutions, biofluids, etc. are a class of the non-Newtonian fluids that demonstrates both the viscous and elastic characters. Stress exhibits here an elastic response to the strain characterized by the relaxation time $\lambda$. Quite intuitively, $\lambda \to 0$ indicates a Newtonian liquid, while the limit $\lambda \to \infty$ stands for the elastic solid. A detailed description of the complex rheological behaviour of these fluids can be found in the monograph by (Bird *et al.* 1987). Here, we are primarily interested in investigating the stability characteristics of a polymeric solution for the most general system configuration: a thin film confined between two bounding surfaces, one rigid (bottom) – other free (top) with an imposed temperature gradient.

The polymeric solutions are the binary mixture of a Newtonian solvent with polymeric solutes having $\lambda \sim \mathcal{O}(10^{-2} - 10^4)$ seconds (Joseph 1990). It is important to note that in presence of a temperature gradient, stratification of the polymeric solutes can take place via the Soret effect (de Gans *et al.* 2003; Würger 2007; Zhang & Müller-Plathe 2006). These solutes, while, usually migrate towards the colder region owing to their large masses, yet, sometimes depending upon the solvent quality and the temperature of the mixture, they can also move to the warmer region. Polymeric solutions thus interestingly demonstrate both the positive and negative Soret effect. Such migration of the solutes can lead to the development of solutocapillary stress on the free surface of the liquid. Hence, Marangoni convection in a polymeric mixture with an imposed temperature gradient arises under the combined influences of thermo- solutocapillarity.

The stability picture for binary mixtures is considerably more complicated than pure liquids. Under the confluence of thermo-solutocapillarity, a binary liquid film gets unstable for heating either from below or above (Castillo & Velarde 1982; Joo 1995; Skarda *et al.* 1998; Podolny *et al.* 2005; Morozov *et al.* 2014). The situation can be expected to get more intricate with the consideration of the elastic behaviour of the mixture. Closer scrutiny of the existing literature in the pertinent paradigm, however, suggests that this binary aspect of a polymeric solution was previously either completely ignored (Getachew & Rosenblat 1985; Dauby *et al.* 1993; Parmentier *et al.* 2000; Hu *et al.* 2016; Lappa & Ferialdi 2018) or the process was analysed by separately considering the thermal and solutal effects (Doumenc *et al.* 2013; Yiantsios *et al.* 2015). Note that a combined thermosolutal model is essential to capture the instability modes that may arise from the interaction between both the driving mechanisms. Nonetheless, these previously reported studies, performed under the assumption of a non-deformable free surface reveal that in a *pure* polymeric film, the disturbances can emerge both in the stationary and oscillatory mode depending upon the level of elasticity of the fluid. The stationary convection arises in a weakly viscoelastic fluid, while the oscillatory instability appears for a highly viscoelastic fluid. The oscillatory instability detected in these works is a sole manifestation of the elastic behaviour of the fluid, which emerges in the short-wave form. Recently, for a *pure* polymeric film confined between its deformable free surface and a poorly conducting substrate, Sarma & Mondal (2019) demonstrated that a long-wave deformational mode can also appear in the system. Notably, in all these previous analyses, the system was considered to get heated from below, leaving the case of heating from above completely unexplored.

These shortcomings motivated us to take up a separate and systematic study on the Marangoni convection in a thin polymeric film subjected to heating from the free surface, considering the binary aspect of the fluid. The primary contribution of this work is to show that the thermo-solutocapillarity, coupled with the elasticity of the fluid, can give rise to two different oscillatory instabilities apart from the monotonic disturbances in this system. The characteristics of each instability mode will be investigated here in detail, identifying the parameter space wherein it can get dominant.

The outline of this paper is the following: in § 2, we describe the physical system considered for investigation and present the governing equations and the related boundary conditions. Identifying the base state, a linear stability analysis is then carried out in § 3. The numerical scheme employed to solve the eigenvalue problem and its validation is briefly discussed in § 4. Our numerical results, presented in § 5, are structured as follows: in § 5.1, we discuss the characteristics of the monotonic mode. The behaviour of the oscillatory disturbances that emerge in two different modes is discussed in §§ 5.2.1 and 5.2.2. The contributions of the fluid elasticity, the thermocapillary and solutocapillary forces towards producing these disturbances are also examined in §5.2. We plot the phase diagrams in § 6 and finally, the main conclusions from this study are summarized in § 7.

## 2. Mathematical formulation

### 2.1. *Governing equations and boundary conditions*

The schematic of the problem under current investigation is illustrated in figure 1. We study the Marangoni instability in a thin layer of an incompressible viscoelastic polymer solution, initially resting on a flat rigid substrate (of lower thermal conductivity compared to the liquid) in the gravitational field **g**. This laterally infinite, two-dimensional film is separated from the ambient gas phase by its deformable free surface located at $z = h(x,t)$. The polymeric solution is a binary mixture of Newtonian solvent and polymeric solute, defined by its relaxation time $\lambda$, viscosity $\mu_o (= \mu_s + \mu_p,\ \mu_s$ and $\mu_p$ are the solvent and solute viscosity, respectively), density $\rho$, thermal conductivity $\kappa$, thermal diffusivity $\alpha$, mass diffusivity $D$ and surface tension $\sigma$.

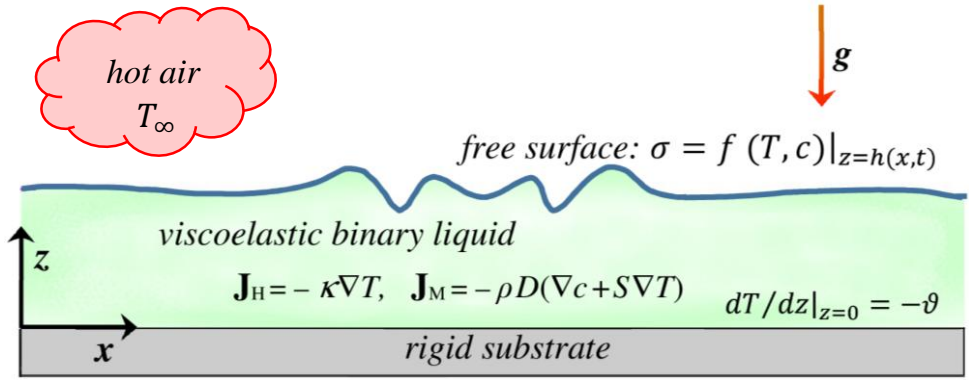

FIGURE 1. (Colour online) Schematic of the physical system under investigation. Marangoni instability is induced in a thin viscoelastic polymer film confined between its deformable free surface (located at $z = h(x,t)$), and a flat substrate (at the $z = 0$ plane) when subjected to heating from above. The polymeric solution is a binary mixture of Newtonian solvent with polymeric solute. The incorporation of the Soret effect signifies combined thermo-solutal instability in the system.

The entire liquid film is subjected to a uniform transverse temperature gradient, specified to be $-\vartheta$ at the $z = 0$ plane. Thus, a negative (positive) $\vartheta$ indicates the case of heating the film from the air-liquid interface (substrate). Here, we are interested in investigating the instability phenomenon only for the former case (i.e. heating the fluid layer at the interface). This applied temperature gradient induces a concentration gradient in the film via the Soret effect. The heat and mass fluxes in the bulk of the fluid layer are thus given by (Groot & Mazur 2011):

$$\mathbf{J}_H = -\kappa \nabla T, \qquad (2.1)$$

$$\mathbf{J}_M = -\rho D (\nabla c + \mathcal{S} \nabla T), \qquad (2.2)$$

respectively, where $T$ is the temperature, $c$ is the solute concentration and $\mathcal{S}$ is the Soret diffusion coefficient of the mixture. As mentioned earlier (see § 1), $\mathcal{S}$ can be either positive or negative for a polymeric mixture depending upon the solvent quality, the mole fractions of the components as well as based on the temperature of the mixture. Note that the Dufour effect through which the concentration gradient couples back to the dynamics of temperature field, is neglected in this analysis owing to its exceedingly weak impact on liquids.

Equations $(2.1) - (2.2)$ indicate that in the conductive state with $H$ as the unperturbed film thickness, the applied heat flux generates a temperature difference $\Delta T = |\vartheta| H$ across the film, which produces a concentration difference $\Delta c = -\mathcal{S} \Delta T$ via the Soret effect. Now, above a certain critical temperature gradient, the thermo- and solutocapillary forces on the free liquid surface induces Marangoni convection in the liquid film. Here, we assume the surface tension to vary monotonically with temperature and concentration of the mixture, dictated by the following relationship

$$\sigma = \sigma_o - \sigma_T (T - T_r) + \sigma_c (c - c_r), \qquad (2.3)$$

where $\sigma_o$ is the surface tension at the reference temperature $T_r$ and concentration $c_r$. For most of the polymer blends $\sigma_T = -\partial \sigma / \partial T |_{T=T_r} > 0$ and $\sigma_c = \partial \sigma / \partial c |_{c=c_r} > 0$ (Doumenc et al. 2013). Here we neglect the effects of buoyancy considering the small thickness of the film ($H \lesssim \mathcal{O}(1)$ mm). Furthermore, except $\sigma$, all other thermo-physical properties are assumed to remain invariant throughout this analysis. Therefore, the evolution of the film velocity $\mathbf{v} \equiv \{u(x,z,t), w(x,z,t)\}$, pressure $p(x,z,t)$, temperature $T(x,z,t)$ and solute concentration $c(x,z,t)$ with time $t$ on the horizontal range $x \in (-\infty, \infty)$ and the vertical range $z \in [0,h]$ are governed by,

$$\nabla \cdot \mathbf{v} = 0, \qquad (2.4)$$

$$\rho \left( \frac{\partial \mathbf{v}}{\partial t} + \mathbf{v} \cdot \nabla \mathbf{v} \right) = -\nabla p + \nabla \cdot \boldsymbol{\tau} - \rho g \mathbf{k}, \qquad (2.5)$$

$$\frac{\partial T}{\partial t} + \mathbf{v} \cdot \nabla T = \alpha \nabla^2 T, \qquad (2.6)$$

$$\frac{\partial c}{\partial t} + \mathbf{v} \cdot \nabla c = D \nabla^2 c + \mathcal{S} D \nabla^2 T, \qquad (2.7)$$

respectively, where $\boldsymbol{\tau} = \begin{bmatrix} \tau_{xx} & \tau_{xz} \\ \tau_{zx} & \tau_{zz} \end{bmatrix}$ is the deviatoric stress tensor, $\boldsymbol{k}$ is the unit vector in the $z$-direction, and $\nabla \equiv \{\partial/\partial x, \partial/\partial z\}$. Note that, the dynamics of the gas and the liquid phases are decoupled here considering the large ratios between their densities, viscosities, and thermal diffusivities.

The boundary conditions that accompany the set of governing equations (2.4)–(2.7) are as follows: at the rigid substrate, we impose the no-slip, no penetration condition for velocity, a specified heat flux, and the mass impermeability conditions, represented respectively by

$$\boldsymbol{v} = \boldsymbol{0}, \quad \frac{\partial T}{\partial z} = -\vartheta, \quad \frac{\partial c}{\partial z} = \mathcal{S}\vartheta \quad \text{at } z = 0. \qquad (2.8a\text{-}c)$$

At the deformable free surface, i.e. at $z = h(x,t)$ the boundary conditions comprise of the kinematic condition

$$w = \frac{\partial h}{\partial t} + u \frac{\partial h}{\partial x}, \qquad (2.9a)$$

that states the velocity of the free surface is equal to the velocity of the liquid, thus giving its location.

The balance of the tangential and normal stress components at the free surface read

$$\frac{1}{\sqrt{1+(\partial h/\partial x)^2}} \left\{ \tau_{xz}\left[1-\left(\frac{\partial h}{\partial x}\right)^2\right] + \tau_{zz}\frac{\partial h}{\partial x} - \tau_{xx}\frac{\partial h}{\partial x}\right\} = \frac{\partial \sigma}{\partial x} + \frac{\partial \sigma}{\partial z}\frac{\partial h}{\partial x}, \qquad (2.9b)$$

$$-p + \frac{1}{1+(\partial h/\partial x)^2}\left[\tau_{zz} + \tau_{xx}\left(\frac{\partial h}{\partial x}\right)^2 - 2\tau_{xz}\frac{\partial h}{\partial x}\right] = \sigma \mathcal{H}, \qquad (2.9c)$$

where $\mathcal{H} = (\partial^2 h/\partial x^2)\left[1+(\partial h/\partial x)^2\right]^{-3/2}$ is the mean curvature.

The thermal boundary condition at the free surface includes the balancing of heat flux across the interface. This heat exchange process with the ambient gas phase is approximated here by the heat transfer coefficient $q$ between the liquid and the gas phase as follows,

$$-\kappa\left(\frac{\partial h}{\partial x}\frac{\partial T}{\partial x} - \frac{\partial T}{\partial z}\right) + q(T-T_\infty)\sqrt{1+(\partial h/\partial x)^2} = 0, \qquad (2.9d)$$

where $T_\infty$ is the uniform gas temperature.

Finally, for this non-volatile binary mixture, the mass flux vanishes at the free surface. Mathematically, this is expressed by

$$\kappa\left(-\frac{\partial h}{\partial x}\frac{\partial c}{\partial x}+\frac{\partial c}{\partial z}\right)-Sq(T-T_\infty)\sqrt{1+(\partial h/\partial x)^2}=0. \qquad (2.9e)$$

## 2.2. *Constitutive equation for the fluid*

Viscoelastic fluids exhibit complex rheology owing to the presence of both the viscous and elastic properties. To depict the rheology of these fluids, a wide variety of constitutive models have been developed over the years. The rheology of the fluid is approximated in this work by the Maxwell model (Bird *et al.* 1987)

$$\boldsymbol{\tau}+\lambda\frac{D\boldsymbol{\tau}}{Dt}=\mu_o\left[(\nabla\boldsymbol{v})+(\nabla\boldsymbol{v})^T\right], \qquad (2.10)$$

which characterize the fluid by a single relaxation time $\lambda$. Here, $\lambda$ is interpreted as the longest relaxation time out of the spectrum of relaxation time exhibited by the liquid. In (2.10),

$$\frac{D\boldsymbol{\tau}}{Dt}=\frac{\partial\boldsymbol{\tau}}{\partial t}+(\boldsymbol{v}\bullet\nabla)\boldsymbol{\tau}-(\nabla\boldsymbol{v})^T\bullet\boldsymbol{\tau}-\boldsymbol{\tau}\bullet(\nabla\boldsymbol{v}) \qquad (2.11)$$

is the upper convected derivative. However, in this investigation, since a linear stability analysis will be carried out for small perturbations around an initially quiescent state, the non-linear terms will not make any contributions to the final results and hence, $D\boldsymbol{\tau}/Dt\equiv\partial\boldsymbol{\tau}/\partial t$. Form (2.10) it is clear that, in the limit $\lambda\to 0$, the Maxwell model depicts the Newtonian fluid behaviour.

## 2.3. *Non-dimensionalization*

Let us now non-dimensionalize the boundary value problem (BVP) formulated by (2.4) – (2.9). Considering the unperturbed film thickness $H$ as the characteristic length scale, the thermal diffusion time $H^2/\alpha$ as the characteristic time scale, and $|\vartheta|H$ as the temperature scale; we define the following set of dimensionless variables:

$$\left.\begin{array}{l}(\bar{x},\bar{z})=\dfrac{(x,z)}{H},\quad \bar{h}=\dfrac{h}{H},\quad \bar{t}=\dfrac{t}{H^2/\alpha},\quad (\bar{u},\bar{w})=\dfrac{u,w}{(\alpha/H)},\quad \bar{\tau}=\dfrac{\tau}{\mu\alpha/H^2},\\[8pt] \bar{p}=\dfrac{p}{\mu\alpha/H^2},\quad \bar{T}=\dfrac{T-T_\infty}{|\vartheta|H},\quad \bar{c}=\dfrac{c}{\sigma_T|\vartheta|H/\sigma_c}.\end{array}\right\} \qquad (2.12)$$

It may be mentioned that the characteristic scale adopted here coincides with the works by Pearson (1958), Shklyaev *et al.* (2009); helping in comparing this work with these previously reported studies. Dropping the over-bar sign from the non-dimensional variables for the convenience in presentation, we finally arrive at the following set of dimensionless governing equations:

$$\nabla \cdot v = 0, \tag{2.13}$$

$$Pr^{-1}\left(\frac{\partial v}{\partial t} + v \cdot \nabla v\right) = -\nabla p + \nabla \cdot \tau - Ga\,k, \tag{2.14}$$

$$\frac{\partial T}{\partial t} + v \cdot \nabla T = \nabla^2 T, \tag{2.15}$$

$$\frac{\partial c}{\partial t} + v \cdot \nabla c = Le\left(\nabla^2 c + \chi \nabla^2 T\right). \tag{2.16}$$

The boundary conditions $(2.8)-(2.9)$ now take the form of

$$v = 0, \quad \frac{\partial T}{\partial z} = -\mathcal{Q}, \quad \frac{\partial c}{\partial z} = \chi \mathcal{Q} \qquad \text{at } z = 0, \tag{2.17a-c}$$

and

$$w = \frac{\partial h}{\partial t} + u\frac{\partial h}{\partial x}, \tag{2.18a}$$

$$\left(\frac{\partial T}{\partial z} - \frac{\partial h}{\partial x}\frac{\partial T}{\partial x}\right) + Bi\,T\sqrt{1+(\partial h/\partial x)^2} = 0, \tag{2.18b}$$

$$\left(\frac{\partial c}{\partial z} - \frac{\partial h}{\partial x}\frac{\partial c}{\partial x}\right) - \chi Bi\,T\sqrt{1+(\partial h/\partial x)^2} = 0, \tag{2.18c}$$

$$-p + \frac{1}{1+(\partial h/\partial x)^2}\left[\tau_{zz} + \tau_{xx}\left(\frac{\partial h}{\partial x}\right)^2 - 2\tau_{xz}\frac{\partial h}{\partial x}\right] = \Sigma\frac{\partial^2 h/\partial x^2}{\left[1+(\partial h/\partial x)^2\right]^{3/2}}, \tag{2.18d}$$

$$\frac{1}{\sqrt{1+(\partial h/\partial x)^2}}\left\{\tau_{xz}\left[1-\left(\frac{\partial h}{\partial x}\right)^2\right] + \tau_{zz}\frac{\partial h}{\partial x} - \tau_{xx}\frac{\partial h}{\partial x}\right\} = Ma\left[\frac{\partial c}{\partial x} - \frac{\partial T}{\partial x} + \left(\frac{\partial c}{\partial z} - \frac{\partial T}{\partial z}\right)\frac{\partial h}{\partial x}\right]$$

$$\text{at } z = h(x,t). \tag{2.18e}$$

Moreover, in non-dimensional form, the Maxwell constitutive model (2.10) reads

$$\tau + De\frac{\partial \tau}{\partial t} = (\nabla v) + (\nabla v)^T. \tag{2.19}$$

Note that, the non-dimensional parameter $\mathcal{Q} = \vartheta/|\vartheta|$ introduced in (2.17) indicates the direction of the applied temperature gradient. $\mathcal{Q} = 1$ represents the case of heating the fluid layer from below, while $\mathcal{Q} = -1$ stands for heating from above. Besides $\mathcal{Q}$, the BVP formulated by $(2.13)-(2.19)$ is further governed by the following set of dimensionless parameters: the Marangoni number, $Ma$, the Soret number, $\chi$, the Deborah number, $De$, the (inverse) Lewis number, *Le*, the Biot number, *Bi*, the Prandtl number, *Pr*, the Galileo number, *Ga*, and the (inverse) capillary number, $\Sigma$:

$$Ma = \frac{\sigma_T |\vartheta| H^2}{\mu_o \alpha}, \quad \chi = \frac{\mathcal{S} \sigma_c}{\sigma_T}, \quad De = \frac{\lambda \alpha}{H^2}, \quad Le = \frac{D}{\alpha},$$

$$Bi = \frac{qH}{\kappa}, \quad Pr = \frac{\mu_o}{\rho \alpha}, \quad Ga = \frac{\rho g H^3}{\mu_o \alpha}, \quad \Sigma = \frac{\sigma H}{\mu_o \alpha}.$$

(2.20)

The Marangoni number governs the present instability phenomenon. *Ma* gives the critical temperature difference across the film ($|\vartheta|H$) at which the convection sets in the film overcoming the stabilizing effects of viscous and thermal diffusion. The Soret number takes into account the relative contributions of the thermocapillary and solutocapillary forces towards the free surface force. Depending upon the Soret coefficient $\mathcal{S}$, $\chi$ can assume both the positive and negative values for a polymeric mixture. The Deborah number quantifies the elastic behaviour of the fluid through the magnitude of $\lambda$. In this analysis, $De = 0$ indicates a Newtonian binary mixture ($\lambda = 0$), while increasing values of *De* signifies enhanced elasticity of the mixture. The (inverse) Lewis number compares the characteristic mass diffusion time scale $H^2/D$ with the thermal diffusion time scale $H^2/\alpha$. The Biot number characterizes the heat transfer rate across the free surface. The Prandtl number is a material property of the fluid that represents the ratio between the thermal diffusion time scale ($H^2/\alpha$) and the viscous diffusion time scale ($\rho H^2/\mu_o$). The Galileo number and the (inverse) capillary number takes account of the deformability of the free surface through the magnitude of **g** and $\sigma$ respectively.

Having formulated the problem, we now conclude this section by discussing the physically permissible limit of the above-mentioned non-dimensional parameters. For a viscoelastic binary mixture, $Pr \gg 1$ and *Le* typically ranges within the interval $\mathcal{O}(10^{-5}) \lesssim Le \lesssim \mathcal{O}(10^{-1})$. Furthermore, we analyse here both the separate cases of $\chi > 0$ and $\chi < 0$, and vary *Bi* within the range $0 < Bi < 1$. To study the stability characteristics of both the weakly and highly viscoelastic fluids, we consider a broad spectrum for *De*: $0 \lesssim De \lesssim \mathcal{O}(10)$. Note that, for a 0.1 mm thick polymeric film with $\alpha \approx \mathcal{O}(10^{-7})$ m$^2$/sec, this range of *De* encompasses fluids having $\lambda \approx \mathcal{O}(0 \sim 10)$ sec.

Moreover, to understand the role of free surface deformations on the onset of instability in the system, here we consider both the cases of deformable and non-deformable free surface. Since the increasing gravitational and surface tension forces reduce the deformability of a free surface, here we consider the microgravity conditions $g \approx \mathcal{O}(0.1)\,\text{m/s}^2$ along with $\sigma \approx \mathcal{O}(10^{-2})\,\text{N/m}$ to depict a deformable free surface. For a 0.1 mm thick layer of the polymeric solution with $\rho \approx \mathcal{O}(10^3)\,\text{kg/m}^3$ and $\mu_o \approx \mathcal{O}(10^{-2})\,\text{Pas}$, the above-mentioned values of $g$ and $\sigma$ yields $Ga = 0.1$ and $\Sigma = 10^3$. On the other hand, the free surface is treated non-deformable in the limit $(Ga, \Sigma) \to \infty$, which typically refers to a liquid layer with significantly high surface tension at the terrestrial environment.

## 3. Base state and linear stability analysis

In the absence of convection in the liquid film, the BVP (2.13)–(2.19) satisfies a no-flow, laterally uniform base state with uniform film thickness. This set of steady solutions is given by

$$\left.\begin{array}{l} \boldsymbol{v}^o = \boldsymbol{0}, \quad \boldsymbol{\tau}^o = \boldsymbol{0}, \quad h^o = 1, \quad p^o = Ga(1-z), \\ T^o = \mathcal{Q}\left(1 - z + Bi^{-1}\right), \quad c^o = \mathcal{Q}\chi z + const. \end{array}\right\} \quad (3.1)$$

In this section, we carry out a linear stability analysis for infinitesimal perturbations around this conductive state of the system. To proceed with, we define the following set of two-dimensional perturbed fields (denoted by a tilde)

$$\left.\begin{array}{l} \boldsymbol{v} = \boldsymbol{v}^o + \tilde{\boldsymbol{v}}(x,z,t), \quad \boldsymbol{\tau} = \boldsymbol{\tau}^o + \tilde{\boldsymbol{\tau}}(x,z,t), \quad p = p^o + \tilde{p}(x,z,t), \\ T = T^o + \tilde{\theta}(x,z,t), \quad h = h^o + \tilde{\xi}(x,z,t), \quad c = c^o + \tilde{c}(x,z,t). \end{array}\right\}. \quad (3.2a-f)$$

Now, substituting these perturbed fields into (2.13)–(2.18), and subsequently linearizing about the base state by obliterating the terms non-linear in perturbations, we finally obtain

$$\nabla \cdot \tilde{\boldsymbol{v}} = 0, \qquad (3.3)$$

$$Pr^{-1} \frac{\partial \tilde{\boldsymbol{v}}}{\partial t} = -\nabla \tilde{p} + \nabla \cdot \tilde{\boldsymbol{\tau}}, \qquad (3.4)$$

$$\frac{\partial \tilde{\theta}}{\partial t} - \mathcal{Q}\tilde{w} = \nabla^2 \tilde{\theta}, \qquad (3.5)$$

$$\frac{\partial \tilde{c}}{\partial t} + \mathcal{Q}\chi\tilde{w} = Le\left(\nabla^2 \tilde{c} + \chi \nabla^2 \tilde{\theta}\right); \qquad (3.6)$$

with the boundary conditions

$$\tilde{\boldsymbol{v}} = \boldsymbol{0}, \qquad \frac{\partial \tilde{\theta}}{\partial z} = 0, \qquad \frac{\partial \tilde{c}}{\partial z} = 0 \qquad \text{at } z = 0, \qquad (3.7a-c)$$

and

$$\frac{\partial \tilde{\xi}}{\partial t} = \tilde{w}, \qquad \frac{\partial \tilde{\theta}}{\partial z} = -Bi\left(\tilde{\theta} - Q\tilde{\xi}\right), \qquad \frac{\partial \tilde{c}}{\partial z} = \chi Bi\left(\tilde{\theta} - Q\tilde{\xi}\right),$$

$$\tilde{\tau}_{xz} = Ma\frac{\partial}{\partial x}\left(\tilde{c} - \tilde{\theta} + Q\tilde{\xi} + Q\chi\tilde{\xi}\right), \qquad -\tilde{p} + Ga\,\tilde{\xi} + \tilde{\tau}_{zz} = \Sigma\frac{\partial^2 \tilde{\xi}}{\partial x^2} \quad \text{at } z = 1,$$

(3.8a–e)

while the constitutive relation (2.19) reads

$$\tilde{\tau} + De\frac{\partial \tilde{\tau}}{\partial t} = \left(\nabla \tilde{v}\right) + \left(\nabla \tilde{v}\right)^T. \tag{3.9}$$

For the sake of convenience, this BVP is now cast in terms of the stream function $\tilde{\psi}(x,z,t)$, such that

$$\tilde{u} = \frac{\partial \tilde{\psi}}{\partial z}, \qquad \tilde{w} = -\frac{\partial \tilde{\psi}}{\partial x}, \tag{3.10a,b}$$

which eliminates the pressure $\tilde{p}$ from the system of equations (3.3)–(3.8). Introducing the stream function relationships (3.10) along with the constitutive equation (3.9), we finally arrive at

$$Pr^{-1}\left(\frac{\partial}{\partial t}\nabla^2\tilde{\psi} + De\frac{\partial^2}{\partial t^2}\nabla^2\tilde{\psi}\right) = \nabla^4\tilde{\psi}, \tag{3.11}$$

$$\frac{\partial \tilde{\theta}}{\partial t} + Q\frac{\partial \tilde{\psi}}{\partial x} = \nabla^2\tilde{\theta}, \tag{3.12}$$

$$\frac{\partial \tilde{c}}{\partial t} - Q\chi\frac{\partial \tilde{\psi}}{\partial x} = Le\left[\nabla^2\tilde{c} + \chi\nabla^2\tilde{\theta}\right], \tag{3.13}$$

with the accompanying boundary conditions,

$$\tilde{\psi} = 0, \qquad \frac{\partial \tilde{\psi}}{\partial z} = 0, \qquad \frac{\partial \tilde{\theta}}{\partial z} = 0, \qquad \frac{\partial \tilde{c}}{\partial z} = 0 \quad \text{at } z = 0, \quad (3.14a–d)$$

$$\frac{\partial \tilde{\xi}}{\partial t} = -\frac{\partial \tilde{\psi}}{\partial x}, \qquad \frac{\partial \tilde{\theta}}{\partial z} = -Bi\left(\tilde{\theta} - Q\tilde{\xi}\right), \qquad \frac{\partial \tilde{c}}{\partial z} = \chi Bi\left(\tilde{\theta} - Q\tilde{\xi}\right), \quad (3.15a–c)$$

$$\frac{\partial^2 \tilde{\psi}}{\partial z^2} - \frac{\partial^2 \tilde{\psi}}{\partial x^2} = Ma\frac{\partial}{\partial x}\left(\tilde{c} - \tilde{\theta} + Q\tilde{\xi} + Q\chi\tilde{\xi}\right) + MaDe\frac{\partial^2}{\partial t\partial x}\left(\tilde{c} - \tilde{\theta} + Q\tilde{\xi} + Q\chi\tilde{\xi}\right), \quad (3.15d)$$

$$\left(1 + De\frac{\partial}{\partial t}\right)\left(\Sigma\frac{\partial^3 \tilde{\xi}}{\partial x^3} - Pr^{-1}\frac{\partial^2 \tilde{\psi}}{\partial t\partial z} - Ga\frac{\partial \tilde{\xi}}{\partial x}\right) = -\frac{\partial}{\partial z}\left(3\frac{\partial^2 \tilde{\psi}}{\partial x^2} + \frac{\partial^2 \tilde{\psi}}{\partial z^2}\right)$$

$$\text{at } z = 1. \quad (3.15e)$$

Here, it is important to note that, since the basic state (3.1) is invariant with respect to $x$ and $t$; we can employ the Fourier decomposition to separate the $x$ and $t$ dependency of the perturbed fields $(\tilde{\psi}, \tilde{\theta}, \tilde{c}, \tilde{\xi})$ from that with $z$:

$$\left(\tilde{\psi}(x,z,t),\ \tilde{\theta}(x,z,t),\ \tilde{c}(x,z,t),\ \tilde{\xi}(x,z,t)\right) = \left(\hat{\psi}(z),\ \hat{\theta}(z),\ \hat{c}(z),\ \hat{\xi}(z)\right)\exp(ikx - \lambda t). \quad (3.16)$$

In (3.16), $(\hat{\psi}, \hat{\theta}, \hat{c}, \hat{\xi})$ represents the amplitude of perturbations, $k$ denotes the dimensionless horizontal wavenumber, and $\lambda = \Omega + i\omega$ refers to the decay rate of the perturbations with $\omega$ (a real quantity) as the perturbation frequency. The dynamics of these infinitesimal perturbations is now governed by the following eigenvalue problem (EVP):

$$Pr\frac{d^4\hat{\psi}}{dz^4} - \left(\lambda^2 De - \lambda + 2Prk^2\right)\frac{d^2\hat{\psi}}{dz^2} + \left(\lambda^2 De - \lambda + Prk^2\right)k^2\hat{\psi} = 0, \quad (3.17)$$

$$\frac{d^2\hat{\theta}}{dz^2} + \left(\lambda - k^2\right)\hat{\theta} = ik\mathcal{Q}\hat{\psi}, \quad (3.18)$$

$$Le\frac{d^2\hat{c}}{dz^2} + \left(\lambda - Lek^2\right)\hat{c} = -Le\chi\left(\frac{d^2\hat{\theta}}{dz^2} - k^2\hat{\theta}\right) - ik\mathcal{Q}\chi\hat{\psi}; \quad (3.19)$$

$$\hat{\psi} = 0, \qquad \frac{d\hat{\psi}}{dz} = 0, \qquad \frac{d\hat{\theta}}{dz} = 0, \qquad \frac{d\hat{c}}{dz} = 0 \quad \text{at } z = 0, \quad (3.20a-d)$$

$$ik\hat{\psi} = \lambda\hat{\xi}, \qquad \frac{d\hat{\theta}}{dz} = -Bi\left(\hat{\theta} - \mathcal{Q}\hat{\xi}\right), \qquad \frac{d\hat{c}}{dz} = Bi\chi\left(\hat{\theta} - \mathcal{Q}\hat{\xi}\right), \quad (3.21a-c)$$

$$\frac{d^2\hat{\psi}}{dz^2} + k^2\hat{\psi} = iMak(1 - \lambda De)\left(\hat{c} - \hat{\theta} + \mathcal{Q}\hat{\xi} + \mathcal{Q}\chi\hat{\xi}\right), \quad (3.21d)$$

$$Pr\frac{d^3\hat{\psi}}{dz^3} + \left(\lambda - \lambda^2 De - 3Prk^2\right)\frac{d\hat{\psi}}{dz} = ikPr(1 - \lambda De)\left(Ga + \Sigma k^2\right)\hat{\xi} \quad \text{at } z = 1. \quad (3.21e)$$

The eigenvalues $\lambda$ and $Ma$ depend on the parameter set ($k$, $Bi$, $De$, $Le$, $\chi$, $Ga$, $\Sigma$, $Pr$) and also on $\mathcal{Q}$.

## 4. Numerical implementation

Solving the EVP for $\mathcal{Q} = -1$ now one can study the stability characteristics of the system for the case of heating from the free surface. However, the complexity of the solvability conditions here restrains us from taking an analytical approach to finding the eigenvalues $\lambda$ and $Ma$. We thus solve equations (3.17)−(3.21) numerically using the fourth-order Runge-Kutta method employing the shooting technique. It is worth mentioning here that the adoption of this technique eliminates the possibility of any spurious eigenvalues, frequently encountered in case of the spectral method (Schmid & Henningson 2001).

The EVP posed by (3.17)–(3.21) suggests the possible emergence of two different instability modes in the system: (i) Monotonic mode (or stationary convection), for this mode, $\lambda = 0$ at the instability threshold; and (ii) Oscillatory mode (or overstability), wherein $\lambda$ attains a purely imaginary value ($= i\omega$) at the instability threshold.

### 4.1. *Validation of the numerical scheme*

Before proceeding further, let us first verify the accuracy of the employed numerical scheme. Note that, although a few studies have been reported in the literature on the Marangoni instability in a (Newtonian) binary liquid film subjected to heating from the free surface, a quantitative comparison with these works cannot be made here due to the adoption of a different system configuration (instead of temperature, its gradient at the solid substrate is specified in the present problem). A qualitative comparison with these works will be provided in the forthcoming sections. The results of the present numerical computation are thus validated against the well-known results of Pearson (1958) (for the monotonic mode in a purely thermocapillary induced instability i.e. for $(\lambda, \chi) = 0$) and Shklyaev *et al*. (2009) (for the oscillatory mode in combined thermo-solutocapillary driven instability) in figure 2.

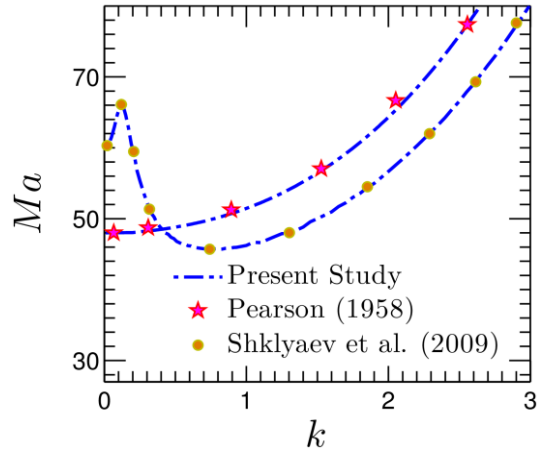

FIGURE 2. (Colour online) Validation of the numerical method: comparison of the present results with the Pearson (1958) and Shklyaev *et al*. (2009) via the neutral stability curve for the monotonic and oscillatory instability mode. To replicate the case of a Newtonian liquid film having an insulated non-deformable free surface subjected to heating from below, we consider $Bi = 0$, $De = 0$, $(Ga, \Sigma) \to \infty$ and $\mathcal{Q} = 1$. For the binary mixture: $Pr = 2$, $\chi = -0.2$ and $Le = 10^{-3}$.

An excellent agreement with the result of the above-cited papers verifies the accuracy of the present numerical implementation.

## 5. Results

Validating the numerical scheme, we now proceed to analyse the stability picture for both the long-wave, $k < \mathcal{O}(1)$ and short-wave, $k \gtrsim \mathcal{O}(1)$ disturbances. Here, we are primarily interested in investigating how viscoelasticity in presence of the Soret effect deviates the stability of the system from its Newtonian counterpart. Let us first start with the monotonic instability mode.

### 5.1. *Monotonic mode*

Figure 3 plots the neutral stability curves $Ma(k)$ for the monotonic instability mode. The solid (dotted) line here refers to a liquid layer with a deformable (non-deformable) free surface. It is observed that, for the entire range of wave-number $k$, there exists a minimum $Ma$ (denoted by the "o" mark) only above which instability sets in the system. We call this $Ma$ as the critical Marangoni number ($Ma_c$), and the corresponding $k$ and $\omega$ as the critical wave number ($k_c$) and oscillation frequency ($\omega_c$), respectively.

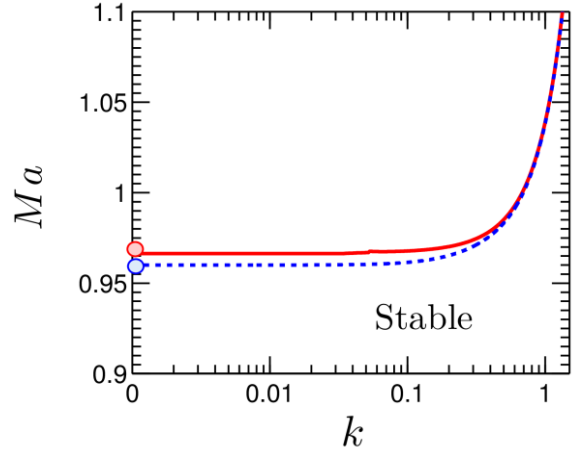

FIGURE 3. (Colour online) Neutral stability curves $Ma(k)$ for the monotonic instability mode at $\chi = -0.5$, $Bi = 0.1$, $De = 1$, $Le = 0.01$, $Pr = 10$. The solid line depicts the stability boundary for a system with deformable free surface $(Ga, \Sigma) = (0.1, 10^3)$; the dotted one shows the instability threshold for a system possessing a non-deformable free surface $(Ga, \Sigma) \to \infty$. The dot (o) mark on each neutral curve represents the critical point of the curve.

Figure 3 shows that for the system subjected to heating from the gas-liquid interface, irrespective of deformability of the free surface, the monotonic disturbances always emerge in the long-wave form ($k_c = 0$). Note that, since the basic state here is simply an origin in the phase space and is identical for both the Newtonian and viscoelastic fluids (i.e. independent of *De*), $Ma_c$ for the onset of stationary convection remains independent of the elasticity of the fluid. In other words, both the Newtonian and viscoelastic binary liquids show identical behaviour towards this particular instability mode. Confirming with the results of Podolny *et al.* (2005), it is found that the increasing deformability of the free surface causes a mild enhancement in the stability of the system against these long-wavelength disturbances, as witnessed in figure 3.

Now, to explore the relative contributions of the thermo- and solutocapillary forces in producing these stationary disturbances, we plot in figure 4 the variation of $Ma_c$ with $\chi$ for both the deformable and non-deformable free surface. Irrespective of deformability of the free surface, the system always remains stable to such disturbances for $\chi \geq 0$, and the instability emerges only when $\chi < 0$. The disappearance of this instability mode for $\chi = 0$, and a reducing $Ma_c$ with $|\chi|$ suggests that the monotonic disturbances are the sole outcome of the solutocapillary effect. The thermocapillarity plays here a stabilizing role. An increasing solutocapillary force for higher values of $|\chi|$ promotes the onset of instability in the system, as can be observed from figure 4.

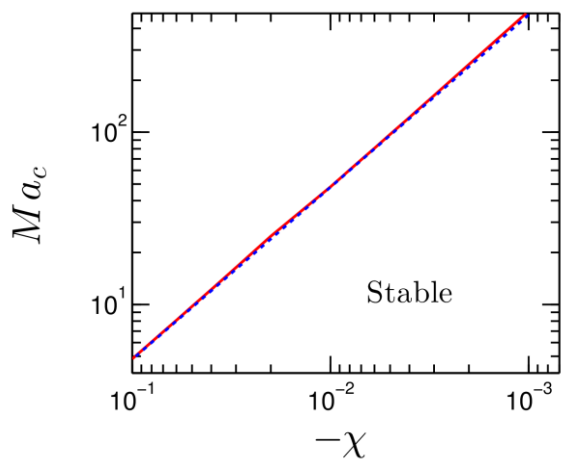

FIGURE 4. (Colour online) Effect of $\chi$ on the stability threshold for the monotonic instability mode at $Bi = 0.01$, $Le = 0.01$, $Pr = 10$. The solid and the dotted line demonstrate the stability boundary for a deformable $(Ga, \Sigma) = (0.1, 10^3)$ and non-deformable $(Ga, \Sigma) \to \infty$ free surface, respectively.

Interestingly, we will demonstrate later on (see § 5.2.2) that, for a shorter mass diffusion time scale $H^2/D$, this counteraction between the two driving forces can give rise to an oscillatory instability in the system.

## 5.2. *Oscillatory mode*

We now study the stability characteristics of the system against the disturbances that emerge with temporal oscillations. Previous investigations (Getachew & Rosenblat 1985; Dauby *et al.* 1993; Parmentier *et al.* 2000) on Marangoni instability in a *pure* viscoelastic film suggest that such a liquid layer is highly vulnerable to this type of disturbances. The present analysis reveals that thermo-solutocapillarity, combined with the elasticity of the fluid, can give rise to two different types of oscillatory instabilities in the system. We call them the oscillatory-I and oscillatory-II mode, respectively. Here, we systematically study the characteristics of both the instability modes by identifying the mechanism behind their inception.

### 5.2.1. *The oscillatory-I instability*

Figure 5 plots the neutral stability curves and the corresponding oscillation frequencies for the oscillatory-I mode. The solid and the dashed line here represents the results for a deformable free surface, while their adjacent dotted line corresponds to a non-deformable free surface for the same *De*. Each neutral curve consists of two branches, both characterized by a distinct local minimum. Note that while one of these local minima lies in the long-wave regime i.e. $k_c < \mathcal{O}(1)$, the other one resides in the short-wave regime $k_c \gtrsim \mathcal{O}(1)$. Accordingly, we call these branches the long-wave and short-wave branch, respectively.

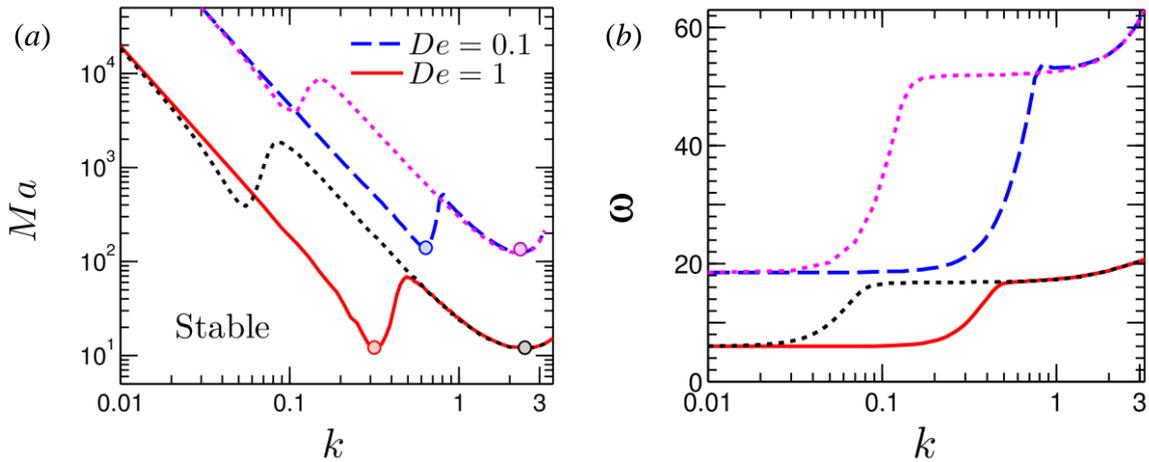

FIGURE 5. (Colour online) (*a*) Neutral stability curves $Ma(k)$ and the corresponding (*b*) oscillation frequency ω for the oscillatory-I mode at $\chi = -0.5$, $Bi = 0.1$, $De = 1$, $Le = 0.01$, $Pr = 10$. The dashed and the solid lines depict the stability boundary for a deformable free surface, while their adjacent dotted line demonstrates the stability threshold for a non-deformable free surface (at the same *De*). The dot (o) mark on each neutral curve represents the critical point (the global minimum) of the curve.

From figure 5(*a*) one can observe that with reducing the deformability of the free surface, $Ma_c$ pertaining to the long-wave branch increases significantly. Thus, for a non-deformable surface, the oscillatory-I disturbance emerges only in the short-wave form. For a given *De*, the parameter set ($Ma_c$, $k_c$, $\omega_c$) for this short-wave branch remains unaffected by the deformability of the free surface. On the other hand, for a liquid layer with a deformable free surface, $Ma_c$ corresponding to the long-wave branch becomes the global minimum. Hence, the first bifurcation occurs here into the long-wave oscillatory-I mode. It is, therefore, evident that a competition between the long-wave and short-wave oscillatory-I disturbances can take place in the system depending upon the deformability of the free surface.

Earlier reported investigations on the Marangoni instability in a Newtonian binary mixture (Skarda *et al.* 1998; Bestehorn & Borcia 2010) suggests that, for heating from above, such a liquid film always remains stable to the oscillatory disturbances if the surface tension reduces with temperature. We now demonstrate that for this particular direction of heating, the oscillatory-I instability can physically emerge only for a highly viscoelastic film and not in a Newtonian or weakly viscoelastic film.

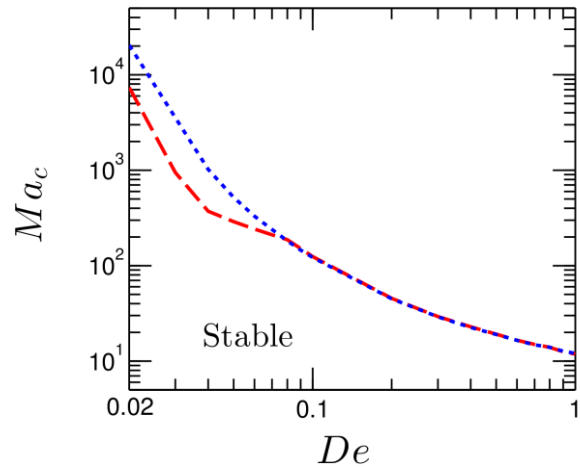

FIGURE 6. (Colour online) Effect of fluid elasticity on the instability threshold for the oscillatory-I mode. The solid and the dotted lines depict the variation for a liquid layer with deformable $(Ga,\Sigma) = (0.1, 10^3)$ and non-deformable free surface $(Ga, \Sigma) \to \infty$, respectively. Other parameters: $Bi = 0.1$, $Le = 0.01$, $Pr = 10$.

From figure 6, it can be observed that for $De = \mathcal{O}(10^{-2})$, $Ma_c \approx \mathcal{O}(10^4)$; while for $De \approx \mathcal{O}(1)$, $Ma_c \approx \mathcal{O}(10)$. Thus, for a 0.1 mm thick film with $\sigma_T \approx \mathcal{O}(10^{-4})$ N/mK, $\mu_o \approx \mathcal{O}(10^{-2})$ Pas, and $\alpha \approx \mathcal{O}(10^{-7})$ m$^2$/sec; the critical temperature difference ($|\vartheta|H$) required to maintain across the film for the onset of oscillatory-I convection is $10^3$ K for $De \approx \mathcal{O}(10^{-2})$; whereas for $De \approx \mathcal{O}(1)$, $(|\vartheta|H) \approx 1$ K. Since $Ma_c$ follows an inverse correlation with $De$, the required temperature difference across the film will be even higher for $De < \mathcal{O}(10^{-2})$, which seems to be completely unrealistic considering the thickness of the film. On the other hand, for $De \gtrsim \mathcal{O}(1)$, $(|\vartheta|H) \lesssim 1$ K, and is experimentally realizable.

*Competition between the thermo- and solutocapillary forces*

Now, to understand the relative contributions of the thermo- and solutocapillary forces in triggering the oscillatory-I disturbances, we plot in figure 7(*a*) the variation of $Ma_c$ with $\chi$.

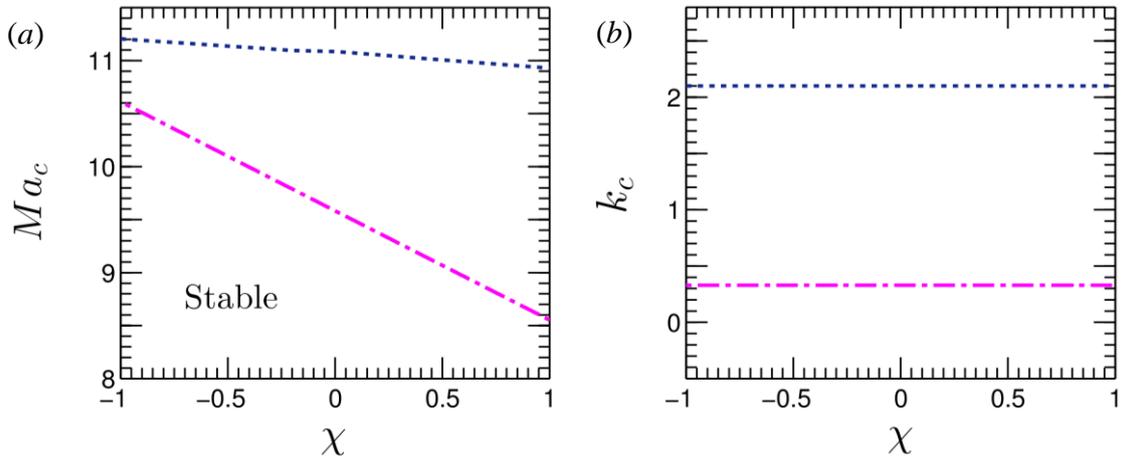

FIGURE 7. (Colour online) Variation of the (*a*) critical Marangoni number $Ma_c$ and the (*b*) critical wave number $k_c$ with $\chi$ for the oscillatory-I instability mode at $Bi = 0.01$, $De = 1$, $Pr = 10$, $Le = 10^{-3}$. In panels (*a,b*), the dash-dotted line represents the results for a deformable $(Ga, \Sigma) = (0.1, 10^3)$, while the dotted line depicts the results for a non-deformable $(Ga, \Sigma) \to \infty$ free surface, respectively.

The emergence of oscillatory-I instability even for $\chi = 0$, and a reducing $Ma_c$ with $De$ (see figure 6) indicates that thermocapillarity, coupled with the elasticity of the fluid, primarily give rise to these disturbances.† The solutocapillary force has only a mild influence in producing this instability mode.

---

†This behavior of the oscillatory-I mode complies with the oscillatory mode found in the case of a *pure* viscoelastic film heated from below. Thus, the oscillatory-I disturbances are essentially the oscillatory disturbance detected therein.

From figure 7(a), it is clear that irrespective of the nature of the Soret effect (i.e. whether $S$ is positive or negative), the oscillatory-I instability can appear for any $\chi \in \mathbb{R}$. However, a key observation here is that for $\chi > 0$, the solutocapillary effect promotes the onset of instability in the system, while it weakly enhances the stability of the system for $\chi < 0$. Interestingly, this contribution of the solutal effect on the stability boundary is affected by the deformability of the free surface. From figure 7(b) one can observe that, for a deformable free surface, the oscillatory-I disturbance emerges in the long-wave form. In contrast, it appears in the short-wave form in case of a non-deformable free surface. Thus, it can be conjectured that, compared to the short-wave disturbances, the solutocapillary effect dominates the stability threshold in a more pronounced manner for the long-wave disturbances.

5.2.2. *The oscillatory-II instability*

In § 5.1, we have mentioned that for this binary mixture subjected to heating from above, an oscillatory instability can emerge in the system only for $\chi < 0$. Clearly this is a different instability mode from the previous one (i.e. the oscillatory-I that can emerge for any $\chi \in \mathbb{R}$). We call this particular mode as the oscillatory-II mode. It will now be demonstrated that the characteristics of this mode are entirely different from the oscillatory-I mode.

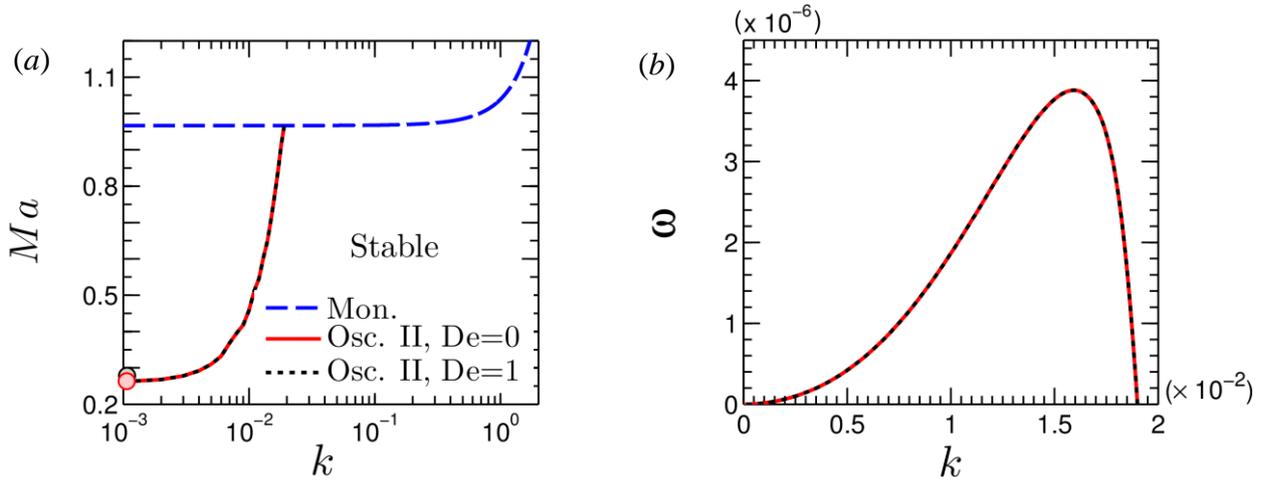

FIGURE 8. (Colour online) (a) Neutral stability curves $Ma(k)$ and the corresponding (b) oscillation frequency ω for the oscillatory-II mode at $\chi = -0.5$, $Bi = 0.1$, $De = 1$, $Le = 0.01$, $Pr = 10$, $(Ga, \Sigma) = (0.1, 10^3)$. The dot (o) mark on each neutral curve in panel (a) represents the critical point of the curve. At higher values of $k$, the neutral curves for the oscillatory-II mode merge with the neutral curves for the monotonic mode.

The typical neutral stability curves and the corresponding oscillation frequency for the oscillatory-II mode is demonstrated in figure 8. It is a long-wavelength instability with $k_c \approx \mathcal{O}(10^{-3})$. At higher values of $k$, the neutral curves for the oscillatory-II mode merge with the neutral curve for the monotonic instability mode. This limits the appearance of these disturbances only in the long-wave form. Another key observation from this figure is that irrespective of the elasticity of the fluid, the oscillatory-II instability can emerge both in the Newtonian and viscoelastic binary liquids with identical $Ma_c$. Hence, *the stability threshold for this particular instability mode remains completely unaffected by the rheological behaviour of the fluid*.

Figure 8(*b*) plots the corresponding oscillation frequencies of the neutral perturbations. As expected, ω for the oscillatory-II mode does not get affected by the viscoelasticity of the fluid. A comparison between figures (5) and (8) further reveals that $k_{c,\text{Osc.-II}} \ll k_{c,\text{Osc.-I}}$ and $\omega_{c,\text{Osc.-II}} \ll \omega_{c,\text{Osc.-I}}$ even for the long-wave oscillatory-I mode. Thus, compared to the oscillatory-I mode, the oscillatory-II disturbances emerge with much larger sized convection cells possessing a higher oscillation period.

*Competition between the thermo- and solutocapillary forces*

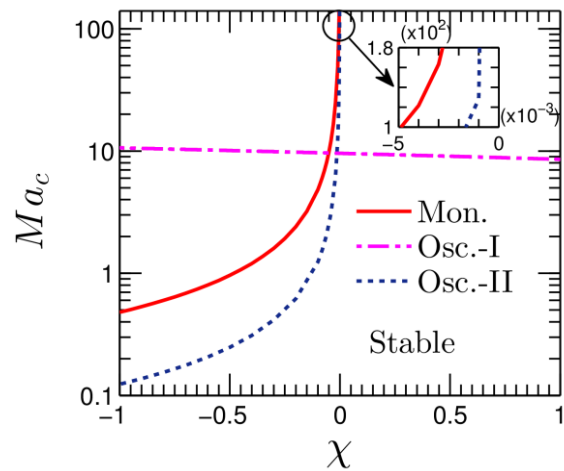

FIGURE 9. (Colour online) Variation of the critical Marangoni number $Ma_c$ with $\chi$ for the oscillatory-II mode (dotted line) at $Bi = 0.01$, $De = 1$, $Le = 0.01$, $Pr = 10$ and $(Ga, \Sigma) = (0.1, 10^3)$. $Ma_c - \chi$ variations for the monotonic (solid line) and the oscillatory-I (dashed-dotted line) modes are plotted here for the reference. Inset shows the zoomed-in view for $\chi \to 0$.

Figure 9 demonstrates that increasing $|\chi|$ promotes the onset of oscillatory-II instability in the system. Note that this particular instability mode emerges only for $Le \gtrsim \mathcal{O}(10^{-2})$ i.e. for a shorter mass diffusion time $(H^2/D)$ compared to the thermal diffusion time $(H^2/\alpha)$. The disappearance of this mode for $\chi = 0$, and a reducing $Ma_c$ with $\chi$ suggests that at the higher rate of solute diffusivity, the increased competition between the destabilizing solutocapillary and the stabilizing thermocapillary forces gives rise to the oscillatory-II disturbances. Similar to the solutocapillarity dominated monotonic mode, this mode, too, always emerges in the long-wave form. However, unlike the former, the oscillatory-II disturbances appear only in the case of a deformable free surface. Although not shown here, these disturbances get damped with reducing the deformability of the free surface, and finally, remain non-existent for a non-deformable free surface.

## 6. Phase diagrams

To get a clear perception of the stability picture, we now plot the phase diagrams in figure 10. The phase diagrams can be helpful in predicting the model parameter space for which a particular instability mode get dominant in the system for bifurcation around the conductive base state. Note that, since the stability characteristics of a viscoelastic binary mixture is studied here under the influence of the Soret effect, the phase diagrams are plotted in a $De - \chi$ plane for different combinations of the parameters $(Le, Ga, \Sigma)$ keeping $Bi$ and $Pr$ fixed. This helps in capturing all possible instability modes that may emerge in the system. In panels (*a-d*) of figure 10, regime-1 represents the monotonic mode, regime-2 the long-wave oscillatory-I mode, regime-3 the short-wave oscillatory-I mode, and regime-4 refers to the oscillatory-II mode. Any dataset corresponding to the boundary between the adjacent instability modes (shown by the dotted line) depicts a competition between them to become the dominant instability mode.

Panel (*a*) plots the phase diagram for a liquid layer with a deformable free surface at $Le = 10^{-3}$. For this system, monotonic disturbances (regime-1) emerge for $\chi < 0$, and the long-wave oscillatory-I instability (regime-2) appears for $\chi \geq 0$.

For a higher rate of the solute diffusivity (i.e. $Le \geq \mathcal{O}(10^{-2})$), panel (*b*) shows that for $\chi < 0$, instead of the monotonic mode, the conductive state first bifurcates into the oscillatory-II mode (regime 4). On the other hand, for $\chi \geq 0$, the long-wave oscillatory-I mode prevails in the system. It is important to note that a competition between the long-wave and short-wave oscillatory-I disturbances can also take place in the system for $\chi \geq 0$ depending upon the deformability of the free surface (see figure 5a).

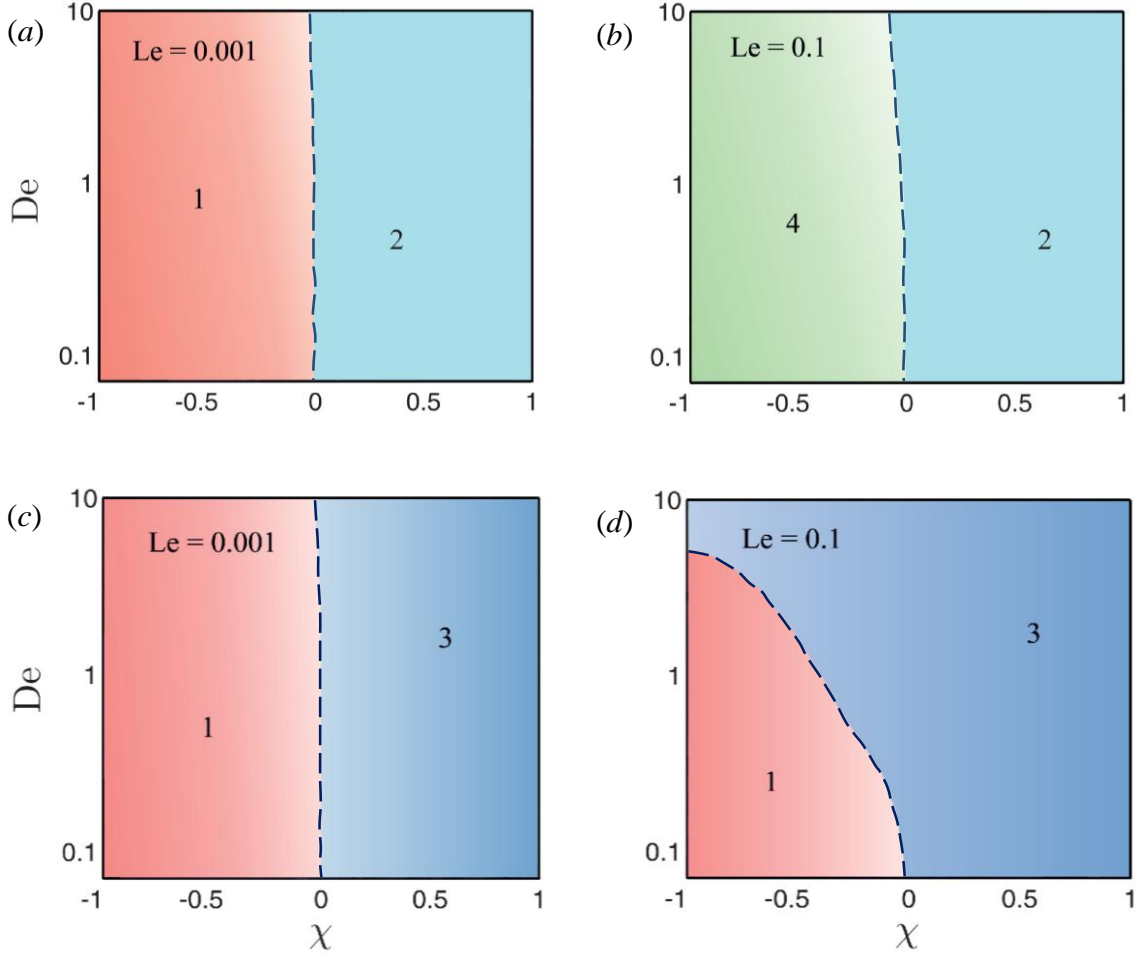

FIGURE 10. (Colour online) Phase diagrams summarizing the boundaries between different dominant instability mode in the $(De, \chi)$ plane for different $Le$: (*a,b*) deformable free surface $(Ga, \Sigma) = (0.1, 10^3)$, and (*c,d*) non-deformable free surface $(Ga, \Sigma) \rightarrow \infty$ at $Bi = 0.01$, $Pr = 10$. In panels (*a-d*), regime-1: monotonic instability, regime-2: long-wave oscillatory–I instability, regime-3: short-wave oscillatory–I instability, regime-4: oscillatory –II instability.

For a non-deformable free surface, panels (*c,d*) show that irrespective of the diffusivity ratio *Le*, the disturbances emerge either in the monotonic or the short-wave oscillatory-I mode. No long-wave oscillatory instability appears here due to dampening out of such disturbances by the increased gravitational and surface tension forces. However, it should be noted that at higher rate of the solute diffusivity (panel *d*), regime-1 shrinks drastically, and thus the conductive state is more likely to lose its stability into the elasticity dominated short-wave oscillatory-I mode.

Together the panels (*a-d*) help in classifying the model parameter space that gives rise to a particular instability mode once the critical temperature difference across the film is attained. Note that, due to uncertainty over data related to the physical properties of the fluid (specifically, $\hbar$ and $\mathcal{S}$ which depend upon the composition of the mixture and demands separate experimentation to determine their values), this critical temperature difference, the corresponding size of the convection cell and its oscillation period at the onset of convection for a realistic system could not be predicted here. Provided a prior estimation of these parameters, the phase diagrams plotted in figure 10 could be helpful for carrying out an experimental investigation of the present problem, especially to study the pattern dynamics for a particular instability mode.

## 7. Concluding remarks

To sum up, here we have conducted, for the first time, a comprehensive study on the Marangoni instability in a viscoelastic binary mixture heated from above by incorporating the Soret effect. The system considered here for investigation comprised of a thin polymeric film confined between its deformable free surface and a rigid substrate. Performing a linear stability analysis around the quiescent base state of this system, we have numerically studied the stability characteristics of the system for both the long-wave and short-wave disturbances. A detailed investigation of the stability picture reveals that apart from the monotonic disturbance, two different oscillatory instabilities, namely the oscillatory-I and the oscillatory-II, can appear in the system under the confluence of thermo-solutocapillary forces and the elasticity of the fluid.

The monotonic mode emerges only for $\chi < 0$, and is solely caused by the solutocapillary effect. The thermocapillarity plays here a stabilizing role, and the instability threshold remains unaltered by the degree of elasticity of the fluid. Irrespective of the deformability of the free surface, such disturbances always emerge in the long-wave form ($k_c = 0$).

The oscillatory-I mode is a direct manifestation of the elastic behaviour of the fluid that can ideally emerge for any $\chi \in \mathbb{R}$. Thermocapillarity, combined with the elasticity of the fluid, primarily gives rise to this instability mode. The solutal effect plays here a minor role in determining the stability threshold. However, the deformability of the free surface plays a crucial role in the emergence of these disturbances in the long-wavelength form, which otherwise appears in the short-wave form for a non-deformable free surface.

The oscillatory-II instability emerges only for $\chi < 0$ on a deformable free surface. For a shorter mass diffusion time scale (i.e. $Le \gtrsim \mathcal{O}(0.01)$), the competition between the destabilizing solutocapillary and the stabilizing thermocapillary forces give rise to these long-wave disturbances. Compared to the oscillatory-I mode, the oscillatory-II disturbances emerge with much larger sized convective structure having a significantly high oscillation period. While the increasing elasticity of the fluid promotes the onset of oscillatory-I disturbances, the oscillatory-II instability threshold essentially remains unaltered by this rheological behaviour.

Thus, the present study establishes that upon heating a polymeric film from the free surface, the solutocapillary force can often destabilize a system which otherwise remains stable under the influence of thermocapillarity. Importantly, under an externally applied temperature gradient, the nonuniformities in solute concentration in the present system are solely caused by the Soret effect. This necessitates the consideration of a complete thermosolutal model to study Marangoni instability in a polymeric film. However, the conditions (i.e. the critical temperature difference across the film, the corresponding size of the convective structure and its oscillation period) at which one may experimentally observe these instability modes for a realistic system is not immediately clear from this study. An estimation of these parameters requires precise knowledge of the rheological properties of the fluid. Given a prior estimation of these parameters, the phase diagrams depicted in figure 10 could be helpful for carrying out an experimental investigation of this problem. We believe the results obtained in this study set an interesting foundation based on which further theoretical and experimental investigations can be carried out to understand the pattern dynamics in the post-critical regime.

## Acknowledgements

Authors acknowledge the computational facilities of Microfluidics and Microscale Transport Processes Laboratory in Mechanical Engineering Department at IIT Guwahati. PKM acknowledges the financial grant obtained from DST-SERB through project No ECR/2016/000702/ES.